\documentclass[aps,twocolumn,prl,showpacs,superscriptaddress,amsmath,amssymb]{revtex4}
\usepackage{graphicx}
\usepackage{dcolumn}
\usepackage{bm}

\newcommand{\figurewidth}{3.2in}

\begin{document}
\bibliographystyle{prsty}

\title{Universality of Ionic Criticality: Size- and Charge-Asymmetric Electrolytes}

\author{Young C. Kim}
\affiliation{Institute for Physical Science and Technology, University of Maryland,
College Park, Maryland 20742}
\author{Michael E. Fisher}
\email[Corresponding author: ]{xpectnil@ipst.umd.edu}
\affiliation{Institute for Physical Science and Technology, University of Maryland,
College Park, Maryland 20742}
\author{Athanassios Z. Panagiotopoulos}
\affiliation{Department of Chemical Engineering, Princeton University, Princeton NJ 08544}

\date{\today}

\begin{abstract}
Grand canonical simulations designed to resolve critical universality classes are reported for $z$:1 hard-core electrolyte models with diameter ratios $\lambda {\,=\,} a_+/a_- {\,\lesssim\,} 6$. For $z {\,=\,} 1$ Ising-type behavior prevails. Unbiased estimates of $T_c(\lambda)$ are within 1\% of previous (biased) estimates but the critical densities are $\sim\,$5 \% lower. Ising character is also established for the 2:1 and 3:1 equisized models, along with critical amplitudes and improved $T_c$ estimates. For $z {\,=\,} 3$, however, strong finite-size effects reduce the confidence level although classical and O$(n {\,\geq\,} 3)$ criticality are excluded.

\end{abstract}
\maketitle

Since the challenging experiments on Coulomb-dominated electrolyte solutions more than a decade ago, the nature of ionic criticality has been an outstanding experimental and theoretical issue \cite{wei:sch}. A central question has been: Is ionic criticality of Ising-type as for simple, nonionic fluids, {\em or} of mean-field character, reflecting the long-range Coulombic interactions, {\em or} something still different? To tackle this problem theoretically, hard-sphere ionic fluids (the so-called {\em primitive models}) have been under broad study analytically and via simulations.

These models consist of $N {\,=\,} N_+ + N_-$ hard sphere ions in a volume $V$, $N_+$ of diameter $a_+$ and charge $q_+ {\,=\,} zq$ and $N_-$ of diameter $a_-$ with charge $q_- {\,=\,} -q$. Ions $\sigma$ and $\tau$ interact via the Coulomb potential $q_\sigma q_\tau/r$. Extensive simulations of the simplest model, namely, the {\em restricted primitive model} (RPM) with $z {\,=\,} 1$ and $a_+ {\,=\,} a_-$, together with appropriate finite-size scaling analyses, have established unequivocally that the critical behavior belongs to the expected $(d {\,=\,} 3)$ Ising universality class with critical exponents $\beta {\,\simeq\,} 0.326$, $\gamma {\,\simeq\,} 1.239$, etc.~\cite{lui:fis:pan,kim:fis:lui,kim:fis}. However, the RPM possesses artificial symmetries of both charge and size. Thus, to better understand ionic criticality, an important question has remained open. Specifically, does the universality class remain unchanged when the symmetry of size or charge is broken? Although experiments on real electrolytes (which are, of course, size-asymmetric) favor Ising-type criticality \cite{wei:sch,kle}, the presence of short-range forces between ions and solvent molecules tends to cloud the theoretical issues regarding the effects of size and charge asymmetry on ionic criticality.

Now long-range Coulombic interactions are known to be screened exponentially at high temperatures and low densities. But whether exponential screening prevails {\em at} criticality is still an open question. How, indeed, does the diverging density correlation length influence charge correlations near criticality? It has been held that the RPM maintains full screening even {\em at} criticality \cite{ste}, thereby supporting expectations of Ising character. On the other hand, Stell \cite{ste} has argued that when size {\em or} charge symmetry is broken, the density fluctuations play a crucial role, mix into the charge correlations, and cause the charge screening length to diverge on approach to criticality. He further suggested that this might change the critical universality class of an ionic fluid leading to mean field behavior \cite{ste}.

Aqua and Fisher \cite{aqu:fis} addressed this issue by analyzing {\em exactly soluble ionic spherical models}. For $z {\,=\,} 1$, they proved that, indeed, when asymmetry is present (which can arise from short-range interactions yielding, in effect, $a_+ {\,\neq\,} a_-$) the density-density and charge-charge correlations mix and the charge correlation length diverges at criticality, as concluded by Stell \cite{ste}. Nevertheless, contrary to Stell's proposal \cite{ste}, the spherical-model criticality is preserved. The coexistence curves of spherical models are, however, always parabolic (i.e., $\beta {\,=\,} \frac{1}{2}$) in contrast to the RPM. Furthermore, spherical models with $z {\,>\,} 1$ have not as yet been studied. Hence it remains a challenge to resolve the question of the critical behavior of asymmetric primitive models.

Some years ago simulations of nonsymmetric primitive models were performed \cite{rom:ork:pan:fis,yan:pab,pan:fis}. It was found that suitably normalized critical temperatures (see below) {\em decrease} with size and charge asymmetry while the critical densities {\em increase} with $z$ but {\em decrease} with size asymmetry. To compute these critical parameters, however, the Bruce-Wilding (BW) method \cite{bru:wil} was employed: this assumes from the outset the Ising character of criticality. Thus, although the simulations were {\em consistent} with Ising behavior, they did not rule out other possibilities, such as classical, i.e., van der Waals (vdW), or selfavoiding walk (SAW), or XY (i.e., $n {\,=\,} 2$), etc. Furthermore, no allowance was made in the analysis for {\em pressure mixing} in the scaling fields \cite{kim:fis2}. 

Our aim here, therefore, is to determine from first principles the universality class of asymmetric primitive models and, further, to obtain {\em unbiased} estimates for the critical parameters taking advantage of recently developed finite-size scaling techniques \cite{kim:fis3}.

Accordingly, we have performed grand canonical Monte Carlo (GCMC) simulations on $z$:1 primitive models. Reduced temperatures and densities are defined naturally by $T^{\ast} {\,=\,} k_{\mbox{\scriptsize B}}T a_{\pm}/zq^{2}$ and $\rho^{\ast} {\,=\,} N a_{\pm}^{3}/V$ where $a_{\pm}=\frac{1}{2}(a_{+}+a_{-})$ is the unlike-ion collision diameter \cite{pan:fis}. The size asymmetry is described by $\lambda {\,=\,} a_+ /a_-$ \cite{rom:ork:pan:fis}. For size asymmetry (with $z {\,=\,} 1$), we compare results for $\lambda {\,=\,} 3$ and $5\frac{2}{3}$ with the RPM $(\lambda {\,=\,} 1)$; for charge asymmetry we study $z {\,=\,} 2$ and 3 with $\lambda {\,=\,} 1$.

Our calculations have been eased by considering finely-discretized primitive models at discretization level $\zeta {\,\equiv\,} a_\pm/a_0 {\,=\,} 10$ so that particles are allowed to sit only on lattice sites of spacing $a_0$ \cite{pan:kum}. Continuum models correspond to the $\zeta {\,=\,} \infty$ limit. However, it has been shown that choosing $\zeta {\,\gtrsim\,} 4$ does {\em not} change the universal critical behavior and makes only small changes in the critical parameters \cite{kim:fis,mog:kim:fis}. Thus there is little doubt that criticality in $(\zeta {\,=\,} 10)$-level models belongs to the same class as in the continuum. To maximize the information gleaned, multihistogram reweighting techniques have been used.

Following our previous strategies for the RPM \cite{lui:fis:pan,kim:fis:lui,kim:fis}, we have studied the {\bf finite-size moment ratios} 
 \begin{equation}
   Q_L(T;\langle\rho\rangle_L) \equiv \langle m^2 \rangle_L^2 /\langle m^4 \rangle_L , \hspace{0.2in} m=\rho-\langle\rho\rangle_L,  \label{eq1}
 \end{equation}
where $\langle\cdot\rangle_L$ denotes a grand-canonical expectation value in a periodic box of side $L$ at fixed $T$ and chemical potential $\mu$ chosen to yield the desired mean density $\langle\rho\rangle_L$. Now, in a single-phase region $Q_L {\,\rightarrow\,} \frac{1}{3}$ when $L{\,\rightarrow\,} \infty$, while $Q_L {\,\rightarrow\,} 1$ on the coexistence-curve diameter. At criticality, $Q_L(T_c,\rho_c)$, approaches a universal value, $Q_c$, that serves to resolve distinct universality classes of criticality rather sharply. Thus, for Ising systems one has $Q_c^{\text{Is}} {\,=\,} 0.623_6$, while the vdW, SAW and XY values are 0.4569$\,\cdots$, 0, and $0.804_5$, respectively; for {\em long-range}, $1/r^{3+\sigma}$ Ising systems with $\frac{3}{2} {\,\leq\,} \sigma {\,\leq\,} 1.96_6$, $Q_c(\sigma)$ rises smoothly from 0.4569$\,\cdots$, reaching $Q_c {\,\simeq\,} 0.600$ at $\sigma{\,=\,} 1.9$ \cite{lui:fis:pan}.

To go further it proves important to calculate $Q_L$ along the $Q$-loci, $\rho_Q(T;L)$, in the $(\rho,T)$ plane defined by the values of $\langle\rho\rangle_L$ for which $Q_L$ is maximal at fixed $T$ \cite{lui:fis:pan,kim:fis3}; at $T {\,=\,} T_c$ these loci approach $\rho_c$ when $L {\,\rightarrow\,} \infty$. Plots of $Q_L(T)$ evaluated on the $Q$-loci for sufficiently large `nearby' sizes, $L {-} 1$ and $L$, cross one another near $T_c$, say at $T_c^{Q}(L)$, with a value, say $Q_c^Q(L)$, close to the corresponding universal value $Q_c$. Finite-size scaling theory \cite{kim:fis3} implies that $Q_c^Q(L)$ approaches $Q_c$ as $L^{-\theta/\nu}$, while $T_c^Q(L)$ approaches $T_c$ more rapidly, as $1/L^{(1+\theta)/\nu}$. For Ising universality, one has $\nu {\,\simeq\,} 0.63$ and $\theta {\,\simeq\,} 0.52$.

\begin{figure}[ht]
\centerline{\includegraphics[width=\figurewidth]{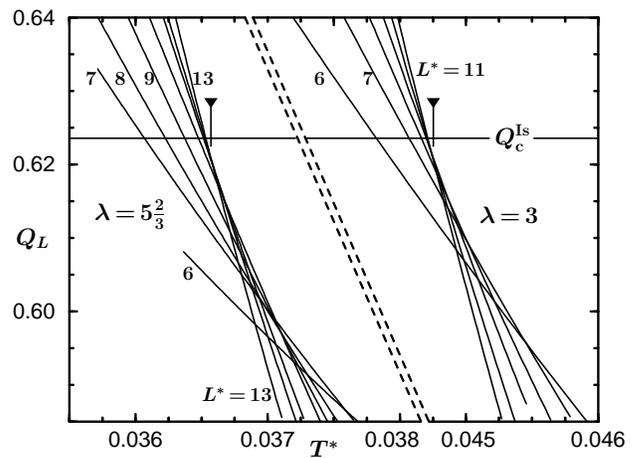}}
\caption{Plots of $Q_L$ on the $Q$-loci vs $T^\ast$ for size asymmetric 1:1 primitive models. The horizontal line marks $Q_L=Q_c^{\text{Is}}$; the arrows locate estimates of $T_c^\ast$.}
\label{fig1}
\end{figure}

Data for {\bf size-asymmetric primitive models} with $\lambda {\,=\,} 3$ and $5\frac{2}{3}$ are shown in Fig.~\ref{fig1} for sizes from $L^\ast {\,\equiv\,} L/a_\pm {\,=\,} 6$ to 13 \cite{note1}. As for the RPM (see Fig.~1 of Ref.~\cite{kim:fis}), the self-intersection points almost coincide with the universal Ising value, $Q_c^{\text{Is}}$. 
Furthermore, $Q_c^{\mbox{\scriptsize vdW}}$ and $Q_c^{\mbox{\scriptsize XY}}$ are off scale while a limit at $Q_c {\,\leq\,} 0.600$ is implausible. Thus these plots not only confirm Ising-type behavior but also exclude vdW, SAW, XY, and long-range Ising criticality with $\sigma\leq 1.9$. Notice, however,that the convergence of $Q_c^Q(L)$ becomes slower when $\lambda$ increases; so a convincing study for larger values of $\lambda$ would require still bigger systems. Nevertheless, we conclude that size asymmetry has no effect on the character of ionic criticality in accord with the spherical model results \cite{aqu:fis}.

On the other hand, all nonuniversal critical parameters depend on $\lambda$ \cite{rom:ork:pan:fis,yan:pab,zuc:fis:bek}. Previous estimates of $T_c$ and $\rho_c$ were obtained via the {\em biased} BW procedure \cite{bru:wil,kim:fis2}; but the plots of Fig.~\ref{fig1} yield the {\em unbiased} precise values presented in Table~\ref{tab1}. These values for $T_c^\ast$ are only 0.2\% lower than the earlier estimates except for a surprising $\sim\,$1\% discrepancy when $\lambda=3$; but our results are more precise by factors of 5 or more.
\begin{table}[ht]
\caption{\label{tab1}Monte Carlo estimates for the size asymmetric 1:1 primitive models at discretization level $\zeta=10$ \cite{pan:kum,kim:fis,mog:kim:fis}. The uncertainties in parentheses refer to the last decimal place.}
\begin{ruledtabular}
\begin{tabular}{c|cc|cc}
  & \multicolumn{2}{c|}{Present work} & \multicolumn{2}{c}{Ref.\ \cite{rom:ork:pan:fis}}  \\
 $\lambda$ & $10^2 T_c^\ast(\lambda)$ & $10^2 \rho_c^\ast(\lambda)~~~~$ & $10^2 T_c^\ast (\lambda)$ & $10^2 \rho_c^\ast (\lambda)~~~~$ \\ \hline
   $~~~~$1$~~~~$   & 4.952(2) & 7.6(2) &  4.96(1) & 7.9(2) \\
   3 & 4.472(2) & 6.1(2)& 4.42(1) & 6.4(4)   \\
   $5\frac{2}{3}$  & 3.654(2) & 4.3(2)& 3.66(8) & 4.6(1) \\
\end{tabular}
\end{ruledtabular}
\end{table}

To obtain the estimates for $\rho_c^\ast$ listed in Table~\ref{tab1} we have, as previously \cite{kim:fis3}, extrapolated the $Q$-loci values, $\rho_Q (T_c^\ast;L)$ vs $1/L^{(1-\alpha)/\nu}$ with $\alpha {\,\simeq\,} 0.11$. The original BW-based estimates are higher than ours by 5, 4 and 7\% (in accord with RPM studies \cite{kim:fis}). These discrepancies may be due to significant pressure mixing \cite{kim:fis2}. Overall, we confirm the previously discovered \cite{rom:ork:pan:fis,yan:pab} strong decrease of $T_c^\ast (\lambda)$ and $\rho_c^\ast (\lambda)$ when $\lambda$ increases.

{\bf The charge-asymmetric primitive models} are of especial interest and pose a greater challenge owing to strong ion-association in the critical region leading to a population of tightly bound neutral and charged clusters \cite{che:pan,aqu:ban:fis}. The models have been studied by simulation \cite{yan:pab,pan:fis,che:pan} and recently an effective Debye-H\"{u}ckel-type theory has been developed \cite{aqu:ban:fis}. One finds that $T_c^\ast (z)$ decreases steadily when $z$ increases while $\rho_c^\ast(z)$ rises sharply \cite{yan:pab,pan:fis,che:pan,aqu:ban:fis}. Although there is a general belief (adopted in the simulations \cite{yan:pab,pan:fis,che:pan}) that increasing $z$ does not alter the Ising critical behavior, there has been no convincing supporting evidence. 

\begin{figure}[ht]
\centerline{\includegraphics[width=\figurewidth]{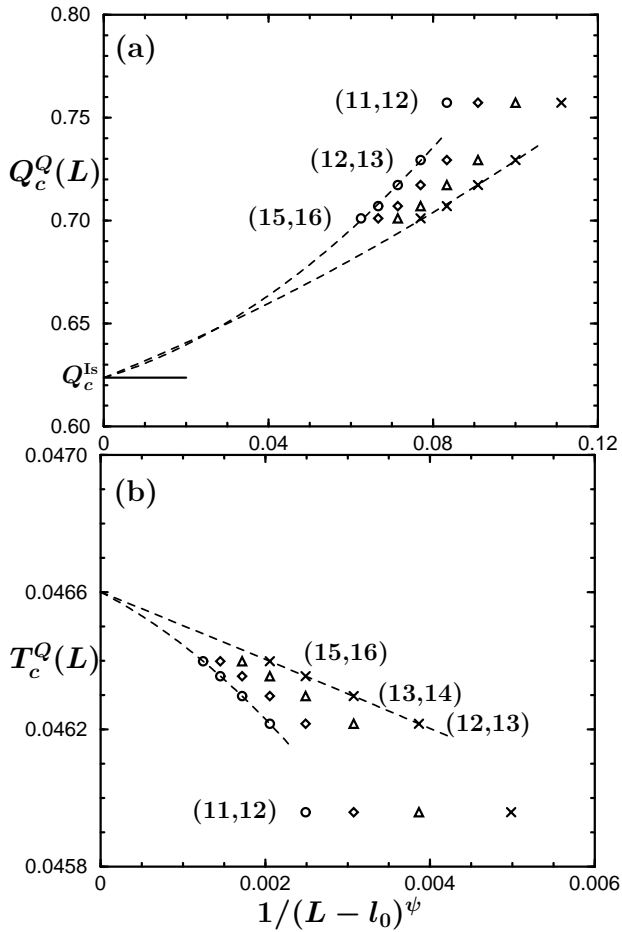}}
\caption{\label{fig2} Plots of $Q_c^Q(L)$ and $T_c^Q(L)$ vs $(L^\ast - l_0)^{-\psi}$ with $\psi {\,=\,} 1$ and $(1+\theta)/\nu {\,\simeq\,} 2.41$, respectively, for the equisized 2:1 electrolyte. The shifts $l_0=0, 1, 2,$ and $3$ (plotted with circles, diamonds, triangles and crosses) are merely aids to extrapolation: the dashed lines are guides to the eye.}
\end{figure}

Our simulations of the equisize 2:1 and 3:1 hard-core models address this issue \cite{note1}. The successive intersections, $Q_c^Q(L)$ and $T_c^Q(L)$, for $z {\,=\,} 2$, found as in Fig.~\ref{fig1}, are presented in Fig.~\ref{fig2} for $L^\ast {\,\geq\,} 11$. 
In contrast to the RPM (see Fig.~\ref{fig1} and Refs.~\cite{lui:fis:pan,kim:fis}), the self-intersections converge much more slowly to their limits as $L$ increases. Nevertheless, the plots in Fig.~\ref{fig2}(a) strongly favor Ising criticality and surely rule out the `adjacent' vdW and XY possibilities at $Q_c {\,\simeq\,} 0.457$ and 0.805. We also find $T_c^\ast(z {\,=\,} 2)=0.0466(1)$ and $\rho_c^\ast(z {\,=\,} 2) =  0.093(3)$; the former is $0.9\%$ lower than the estimate 0.047, obtained by using the BW method {\em at} $L^\ast {\,=\,} 15$ \cite{pan:fis}, while the value for $\rho_c^\ast$ agrees \cite{pan:fis}. 

\begin{figure}[ht]
\centerline{\includegraphics[width=\figurewidth]{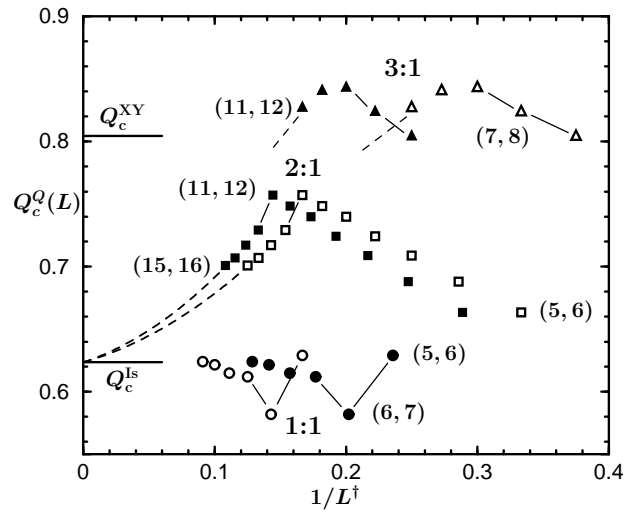}}
\caption{\label{fig3} Plots of $Q_c^Q(L)$ vs $1/L^\dagger$ with $L^\dagger \equiv L^\ast/z$ (open symbols) and $L^\ast/\sqrt{z+1}$ (solid symbols) for $z {\,=\,} 1$, 2 and 3.}
\end{figure}

To understand the behavior of $Q_c^Q(L)$ in further detail, we also simulated smaller systems, $L^\ast {\,\leq\,} 12$, for $z {\,=\,} 2$ and $3$ \cite{note2}. Surprisingly, the behavior of the $Q$-intersections turns out to be sharply nonmonotonic in $L$: see Fig.~\ref{fig3}. In fact, this is also seen for the RPM $(z {\,=\,} 1)$ when data for $L^\ast {\,\leq\,} 6$ are examined. Although unexpected, the behavior is clearly a reflection of strong finite-size effects. To gain a feeling for the phenomenon, note that in GCMC simulations, one attempts to insert a {\em neutral} cluster or ``molecule'' of $z+1$ ions. When $z$ increases, larger systems are evidently needed to accommodate the same number of molecules at comparable {\em molecular} densities. One may thus enquire into the geometry of neutral ion clusters. To reflect this, the data in Fig.~\ref{fig3} have been plotted using a system size, $L$, rescaled by $z a_\pm$ (open symbols) and by $\sqrt{z+1}\, a_\pm$ (solid symbols). The first scaling embodies the center-to-center size of a neutral cluster on a line; indeed, for $z {\,=\,} 1$ and 2 the molecular ground states are linear. But for $z {\,=\,} 3$ the ground state is a planar centered triangle \cite{aqu:ban:fis}: the second rescaling respects this feature and seems more appropriate.

Also relevant and seen in Fig.~\ref{fig3} is the sharp changeover in behavior with a strong $z$-dependence at relatively small sizes. Finite-size scaling theory \cite{kim:fis3} predicts $[Q_L\bm{(}T_c^Q(L)\bm{)}-Q_c] \sim L^{-\theta/\nu}$ with higher order terms $L^{-2\beta/\nu}$, $L^{-(1-\alpha)/\nu}$, etc. The interference of these terms may certainly lead to nonmontonic behavior; however, naive fits to the data with a few leading terms are {\em not} successful. Furthermore, with only sizes up to $L^\ast {\,=\,} 11$ available one would be led to guess that the limiting $Q_c$ for $z {\,=\,} 2$ and 3 exceeded 0.8 or 0.9, respectively; but, as clear from the $z {\,=\,} 2$ data, this is not sustainable. Nevertheless, on the basis of Fig.~\ref{fig3} it is tempting to speculate that the $z {\,\geq\,} 2$ equisized primitive models may exhibit some sort of crossover from $n {\,=\,} \infty$ behavior (with $Q_c^\infty {\,\simeq\,} 1.0$) for $z {\,\gg\,} 1$ to $n {\,=\,} 1$, Ising behavior for finite $z$ as $L$ increases. But whether this is a genuine effect surely needs further investigation!

In the absence of precise data for significantly larger systems \cite{note2}, we cannot convincingly argue that the 3:1 equisized model exhibits Ising behavior. Thus, while vdW criticality seems excluded, the turnover in Fig.~\ref{fig3} for  $L^\ast {\,=\,} 12$, is {\em consistent} with XY or O$(n {\,=\,} 2)$ criticality. Nevertheless, by comparison with the $z {\,=\,} 2$ data (which extend to $L^\ast {\,=\,} 16$), our belief that Ising criticality will still be realized when $L^\ast {\,\rightarrow\,} \infty$ is surely reasonable. Accepting that, one may estimate $T_c^\ast$ by extrapolating the intercepts of $Q_L(T^\ast)$ with $Q_c^{\text{Is}}$; then $\rho_c^\ast (z{\,=\,} 3)$ follows from the $Q$-loci, as for $z {\,=\,} 1$ and 2. This yields $T_c^\ast (z {\,=\,} 3) {\,=\,} 0.0405(2)$ and $\rho_c^\ast (z {\,=\,} 3) {\,=\,} 0.126(3)$ which may be compared with 0.0410(1) and 0.125(4), respectively, obtained by the BW procedure at $L^\ast {\,=\,} 15$ \cite{pan:fis}.

Finally, the critical amplitudes are of interest \cite{fis:zin}. For the RPM we have $C^+ a_\pm^3 {\,\simeq\,} 0.087$ for the susceptibility and $B a_\pm^3 {\,\simeq\,} 0.284$ for the order parameter \cite{kim}. Together with the universal ratio $Q_c {\,\equiv\,} (\xi_{1,0}^+)^d B^2/C^+ {\,\simeq\,} 0.323$ \cite{fis:zin} these yield the density correlation-length amplitude, $\xi_{1,0}^+/a_\pm {\,\simeq\,} 0.704$; this is surprisingly close to the generalized-Debye-H\"{u}ckel (GDH) value 0.7329 \cite{lee:fis}. For $z {\,=\,} 2$ and 3, we now find, to lower precision, $C^+ a_\pm^3 {\,=\,} 0.11(1)$, 0.16(3) and $B a_\pm^3 {\,=\,} 0.35(5)$, 0.45(5) \cite{pan:fis}, respectively, which yield $\xi_{1,0}^+/a_\pm {\,=\,} 0.66(5)$ and 0.63(9), not inconsistent with the RPM value. The relative insensitivity of $\xi_{1,0}^+$ to $z$ is notable.

In summary, we have studied the critical behavior of {\em size}- and {\em charge-asymmetric} primitive model electrolytes via extensive Monte Carlo simulations. It has been shown convincingly for $z {\,\leq\,} 2$ that breaking either symmetry leaves the universal Ising character unchanged. For the equisized charge-asymmetric 3:1 electrolyte, strong finite-size effects mandate simulations of much larger systems; nevertheless, the data are consistent with Ising-type behavior and van der Waals and O$(n {\,\geq\,} 3)$ criticality are excluded. The density correlation-length amplitudes vary little with $z$ and are to be close to the prediction of GDH theory \cite{lee:fis}.

We are indebted to Erik Luijten for his histogram reweighting program and to Jean-No\"{e}l Aqua for valuable discussions. The support of the National Science Foundation (via Grant No.\ CHE 03-01101 to M.\ E.\ F.) and the Department of Energy (Grant DE-FG201ER15121 to A.\ Z.\ P.) is gratefully acknowledged.


\begin{thebibliography}{99}
\bibitem{note2}
As $z$ increases, the acceptance ratio decreases dramatically, making reliable simulations of larger systems for $z {\,=\,} 3$ computationally impractical. To go significantly further, new techniques need to be explored.
\end{thebibliography}

\end{document}